\documentclass[preprint,aps,pra,showpacs,floatfix,superscriptaddress]{revtex4-1}
\usepackage{graphicx}
\usepackage{feynmp}
\usepackage{amsmath,amsfonts,amssymb,verbatim,float}
\usepackage{epsf}
\usepackage{bm}
\usepackage{bbm}
\usepackage[usenames]{color}	%for colours
\usepackage{colortbl}
\usepackage{hyperref} % for links in the bibliography
\usepackage{indentfirst}
\usepackage{siunitx} % vyravnivanie po tichke
\usepackage{calrsfs}
\usepackage{calrsfs}
\usepackage{ulem} % for remarks
\usepackage{subfigure}
\newcommand{\br}{{\bf r}}

\newcommand{\balpha}{\boldsymbol{\alpha}}

\graphicspath{{graph/}}
\newcommand{\be}{\begin{eqnarray}}
\newcommand{\ee}{\end{eqnarray}}
\newcommand{\la}{\langle}
\newcommand{\ra}{\rangle}

\newcommand{\veps}{\varepsilon}

\newcommand{\pr}{\prime}

\newcommand{\bmu}{\bm{\mu}}
\newcommand{\bfT}{{\bf T}}

\newcommand{\bfn}{{\bf n}}

\newcommand{\bfr}{{\bf r}}

\newcommand{\aZ}{\alpha Z}

\newcommand{\ket}[1]{|#1\rangle}
\newcommand{\braket}[2]{\langle#1|#2\rangle}
\newcommand{\matrixel}[3]{\langle #1 | #2 | #3 \rangle}

\newcommand{\intG}{\oint_{\Gamma}}
\newcommand{\tomega}{\tilde\omega}
\begin{document}
\thispagestyle{empty}
\title{
Many-electron effects in the hyperfine splitting of lithiumlike ions
}
\author{V.~P.~Kosheleva}
\affiliation{
Helmholtz Institute Jena, 07743 Jena, Germany
}
\affiliation{
GSI Helmholtzzentrum f\"ur Schwerionenforschung GmbH, 64291 Darmstadt, Germany
}
\affiliation{
Theoretisch-Physicalisches Institut, Friedrich-Schiller-Universit\"at, 07743 Jena, Germany
}
\affiliation{
Department of Physics, St. Petersburg State University, 199034 St. Petersburg, Russia
}
\author{A.~V.~Volotka}
\affiliation{
Helmholtz Institute Jena, 07743 Jena, Germany
}
\affiliation{
GSI Helmholtzzentrum f\"ur Schwerionenforschung GmbH, 64291 Darmstadt, Germany
}
\author{D.~A.~Glazov}
\affiliation{
Department of Physics, St. Petersburg State University, 199034 St. Petersburg, Russia
}
\author{S.~Fritzsche}
\affiliation{
Helmholtz Institute Jena, 07743 Jena, Germany
}
\affiliation{
GSI Helmholtzzentrum f\"ur Schwerionenforschung GmbH, 64291 Darmstadt, Germany
}
\affiliation{
Theoretisch-Physicalisches Institut, Friedrich-Schiller-Universit\"at, 07743 Jena, Germany
}
%
%======================       ABSTRACT         =========================
\begin{abstract}
The rigorous QED evaluation of the one- and two-photon exchange corrections to the ground-state hyperfine splitting in Li-like ions is presented for the wide range of nuclear charge number $Z= 7 - 82$.
The calculations are carried out in the framework of the extended Furry picture, i.e., with inclusion of the effective local screening potential in the zeroth-order approximation.
The interelectronic-interaction contributions of the third and higher orders are taken into account in the framework of the Breit approximation employing the recursive perturbation theory. 
In comparison to the previous theoretical calculations, the accuracy of the interelectronic-interaction contributions to the ground-state hyperfine splitting in Li-like ions is substantially improved.
\end{abstract}
%======================    END ABSTRACT       ==========================
%
%%%%%%%%%%%%%%%%%%%%%%%%%%%%%%%%%%%%%%%%%%%%%%%%%%%%%%%%%%%%%%%%%%%%%%%%
\maketitle
%%%%%%%%%%%%%%%%%%%%%%%%%%%%%%%%%%%%%%%%%%%%%%%%%%%%%%%%%%%%%%%%%%%%%%%%
%
% ======================== INTRODUCTION ================================
%
\section{Introduction}
Investigations of the hyperfine structure (hfs) in highly charged, few-electron ions, triggered by the first measurements in H-like ions \cite{klaft:1994:2425, crespo:1996:826, crespo:1998:879, seelig:1998:4824, beiersdorfer:2001:032506}, provide unique possibilities to test QED in the strongest electric and magnetic fields. 
Present theoretical studies are motivated by the experimental breakthrough in measuring the hfs in H- and Li-like bismuth ions which, nowadays, has reached an accuracy of less than $0.002\%$ \cite{lochmann:2014:R030501, ullmann:2015:144022, sanchez:2017:085004, ullmann:2017:15484}.
Such precise measurements of hfs in both H- and Li-like ions of the same isotope allow probing QED in the strong-field regime within the concept of the specific difference \cite{shabaev:2001:3959}, which, however, reveals the $7\sigma$ discrepancy between experimental \cite{ullmann:2017:15484} and theoretical \cite{volotka:2012:073001} values of the specific difference in $^{209}$Bi.
The reason for this discrepancy has meanwhile been explained by the incorrect value of the nuclear magnetic moment of $^{209}$Bi. A new value of the magnetic moment, obtained in the recent NMR experiment together with the elaborated magnetic shielding calculations, strongly differs from the tabulated one \cite{skripnikov:2018:093001}. Although the current value brings a specific difference into agreement, its significantly larger uncertainty limits the test of QED in $^{209}$Bi. In particular, the uncertainty due to the magnetic moment is about one order of magnitude larger than other uncertainties in the theoretical value of the specific difference \cite{noertershaeuser:2019:240}. To push forward the test of QED with the hfs we extend the calculations to other Li-like ions.
\\
\indent
Various QED and interelectronic-interaction contributions to the ground-state hfs in few-electron ions were investigated in past decades. The leading one-electron QED corrections to the hfs due to the one-loop self-energy and vacuum polarization diagrams were calculated earlier for $1s$ \cite{schneider:1994:118, persson:1996:204, blundell:1997:1857, shabaev:1997:252, sunnergren:1998:1055, yerokhin:2001:012506, yerokhin:2005:052510, volotka:2008:062507, yerokhin:2010:012502}, $2s$ \cite{yerokhin:2001:012506, yerokhin:2005:052510, volotka:2008:062507, yerokhin:2010:012502, shabaev:1998:149, sapirstein:2001:032506, oreshkina:2007:889}, $2p_{1/2}$ \cite{yerokhin:2010:012502, sapirstein:2006:042513, oreshkina:2008:675, glazov:2019:062503, volotka:2008:062507}, and $2p_{3/2}$ \cite{yerokhin:2010:012502, sapirstein:2008:022515} states for H-like ions as well as for Li- and B-like ions and more recently for many-electron neutral atoms \cite{sapirstein:2003:022512, ginges:2017:062502}.
For many-electron systems, the QED contributions were evaluated by using local screening potential approximation -- the so-called extended Furry picture, which implies an additional effective local screening potential in the zero-order Hamiltonian.
Lately, the two-electron self-energy \cite{volotka:2009:033005, glazov:2010:062112} and a major part of the two-electron vacuum-polarization \cite{andreev:2012:022510} corrections were calculated in the framework of the QED approach which yields the correct result to all orders in $\alpha Z$ ($\alpha$ is the fine structure constant, $Z$ is the nuclear charge number).
The existing theoretical calculations of electronic correlations are based on the multiconfiguration Dirac-Fock \cite{boucard:2000:59}, all-orders correlation potential \cite{dzuba:1987:1399, ginges:2018:032504}, configuration interaction \cite{shabaev:1995:3686, zherebtsov:2000:701, yerokhin:2008:R020501}, many-body \cite{blundell:1989:2233} or QED perturbation theory methods. Since the QED formalism does not allow to incorporate interelectronic-interaction effects to all orders in $1/Z$, it is usually merged with methods based on the Breit approximation. The one-photon exchange correction, which refers to the first order in $1/Z$ was firstly derived in Ref. \cite{shabaeva:1995:2811} and nowadays is routinely calculated.
The rigorous evaluation of the two-photon exchange correction to the ground-state hfs in Li-like $^{209}$Bi$^{80+}$ was presented in Ref.~\cite{volotka:2012:073001}. The calculation was done in the framework of the original Furry picture, i. e., without effective screening potential, the higher-order interelectronic-interaction contributions were taken into account within the large-scale configuration-interaction Dirac-Fock-Sturm method (CI-DFS) \cite{bratsev:1977:2655}.
\\
\indent
In the present paper we report a complete evaluation of the one- and two-photon exchange corrections to the hfs in Li-like ions for a wide range of nuclear charge $Z$ within the framework of rigorous QED approach in the extended Furry picture. The higher-order interelectronic-interaction contributions have been taken into account through the recursive perturbation theory \cite{glazov:2017:46}.  
In the framework of the extended Furry picture, the interelectronic interaction is partly taken into account already in the zeroth order what allows to accelerate the convergence of perturbation theory. 
As a result, we substantially improve the accuracy of the interelectronic-interaction contribution to the hfs in Li-like ions through a wide range of the nuclear charge number $Z = 7 - 82$.
\\
\indent
The paper is organized as follows. In Section~\ref{sec-1} the basic formalism for the ground-state hfs in Li-like ions is given.
In Section~\ref{sec-2} we present the consistent evaluation of the interelectronic-interaction corrections to the hfs in Li-like ions of the first (\ref{subsec-B}), second (\ref{subsec-C}), and higher orders (\ref{subsec-D}).
Finally, in Section~\ref{sec-3} we report the numerical results obtained as well as the comparison of the QED treatment of the interelectronic interaction with methods based on the Breit approximation.
\\
\indent
Relativistic units ($\hbar = 1$, $c = 1$, $m_e = 1$) and the Heaviside charge unit [$\alpha = e^2/(4\pi)$, $e<0$] are used throughout the paper.
%
% ======================== BASIC FORMALISM =============================
%
\section{Basic formulas}
\label{sec-1}
The hyperfine splitting of atomic energy levels arises from the interaction of bound electrons with the magnetic field of the nucleus. 
In the dipole approximation this interaction is described by the Fermi-Breit operator: 
\be
  H_\mu = \frac{|e|}{4\pi}\,\bmu \cdot \bfT\,,
  \label{eq:Hhfs} 
\ee
where $\bmu$ is the nuclear magnetic moment operator acting in the space of nuclear states. The electron part $\bfT$ is defined by the following expression:
\be
\label{bfT}  
  \bfT = \sum_i \frac{[\bfn_i\times\balpha_i]}{r_i^2}F(r_i)\,.
\ee
Here index $i$ refers to the $i$th electron of the atom, $\balpha$ is the Dirac-matrix vector, $\bfn_i = \bfr_i/r_i$, and $F(r_i)$ is the nuclear-magnetization volume distribution function discussed below.
Due to this interaction, the angular momentum of atomic electrons ${\bf J}$ and the nuclear spin ${\bf I}$ are not conserved separately, and only the total atomic angular momentum ${\bf F} = {\bf J} + {\bf I}$ is an integral of motion. Therefore the energy levels, characterized by quantum number $J$, split into sublevels corresponding to all possible values of the total angular momentum $F$:
\begin{equation}
 F = J+I,J+I-1, \cdots, \mid J - I \mid\,.
 \label{eq:F} 
\end{equation}
This splitting is known as the hyperfine splitting. Here, we restrict our consideration to the Li-like ions  with the valence electron in a state $\vert a \rangle = \vert j_a m_a \rangle$ with total angular momentum $j_a = 1/2$ and its projection $m_{a}$. In this case, the angular quantum numbers of the electronic system are determined by the valence electron $\vert a \rangle$: $J = j_a$ and $M_J = m_a$, with $M_J$ being the projection of $J$. Then according to Eq.~\eqref{eq:F} the energy levels of Li-like ion split into two components: $F^{+} = I+1/2$ and $F^{-} = I-1/2$ and the ground-state hfs value in Li-like ions can be written as follows:
\begin{equation}
\Delta E_{\rm hfs}  = E(F^{+}) - E(F^{-})\,,
\label{hfs1}
\end{equation}
where $E(F)$ is the energy level of the Li-like ion with the total angular momentum $F$. In the nonrelativistic one-electron point-nucleus approximation the ground-state hfs can be calculated analytically (so-called Fermi energy $E_{\rm F}$):
\begin{equation}
\Delta E_{\rm hfs} \underset{\rm nonrel}{\longrightarrow} E_{\rm F} 
= \frac{\alpha(\aZ)^3}{n_a^3} \frac{g_I}{m_p}
    \frac{2I+1}{(j_a+1)(2l_a+1)}\frac{1}{(1+\frac{m_e}{M})^3}\,.
\label{hfs2}
\end{equation}
Here $l_a = j_a \pm 1/2$ defines the parity of the state $\vert a \rangle$, $n_a$ is the principal quantum number of a valence electron, $g_I = \dfrac{\mu}{\mu_N I}$ is the nuclear $g$ factor, $\mu$ is the nuclear magnetic moment, and $\mu_N = \dfrac{|e|}{2 m_p}$ is the nuclear magneton, $m_e$, $m_p$, and $M$ are the electron, proton, and nuclear masses, respectively.
\\
\indent
Using Eq.~\eqref{hfs2} it is convenient to introduce the following parametrization of the ground-state hfs in Li-like ion:
\begin{equation}
\Delta E_{\rm hfs} = E_{\rm F}X_{a}(1 - \epsilon)\,,
\label{hfs3}
\end{equation}
where $\epsilon$ is a correction due to the spatial distribution of the nuclear magnetization, the Bohr-Weisskopf correction, and $X_{a}$ is a dimensionless hfs factor incorporating the many-electron and QED effects.
\\
\indent
The Bohr-Weisskopf correction $\epsilon$ originates from the extended nuclear magnetization distribution can be taken into account by the volume distribution function $F(r)$ in Eq.~\eqref{bfT}, which is equal to one, $F(r) = 1$, for the point-like nuclear magnetic moment. In the present work we use the homogeneous sphere model, for this case $F(r)$ reads as follows:
\be
\label{BW:2}
  F(r) = \left\{\begin{array}{cr}
     \displaystyle\left(\frac{r}{R_0}\right)^3, & r \le R_0\\
     \displaystyle                           1, & r > R_0
                \end{array}\right.\,,
\ee
where $R_0 = \sqrt{\dfrac{5}{3} \langle r^{2}\rangle}$ is the radius of the magnetization sphere and the corresponding root-mean-square (rms) radius $\langle r^{2}\rangle$ is assumed to be the same as the nuclear charge rms radius. We will discuss the choice of the nuclear model in Section \ref{sec-3}.
\\
\indent 
The dimensionless hfs parameter $X_{a}$ in the one-electron approximation is given by:
\begin{equation} 
\label{K_0}
 X_{a}= G_a \langle  a \vert  T_0 \vert  a \rangle\,,
\end{equation}
with
\be
 G_a = \frac{n_a^3 (2l_a+1) j_a (j_a+1)}{2 (\aZ)^3 m_{a}}\,.
\ee
Here $T_0$ refers to the zeroth component of the electron part of the Fermi-Breit operator given by Eq.~\eqref{bfT}. $G_a^{-1}$ is the nonrelativistic value of the electron part of the Fermi energy, so that $X_a \longrightarrow 1$ in the nonrelativistic one-electron point-nucleus approximation.
%
% ======================== MANY-ELECTRON EFFECTS =======================
%
\section{Many-electron effects}
\label{sec-2}
Let us now consider the many-electron effects to the hfs in the framework of the QED perturbation theory.
The interaction Hamiltonian $H_I$ can be written as a sum $H_I = H_{\rm QED} + H_{\mu}$, where $H_{\rm QED}$ is the usual QED Hamiltonian \cite{mohr:1998:227} describing the interaction between the electron-positron field and the photon field and $H_{\mu}$ is the Fermi-Breit operator, see Eq.~\eqref{eq:Hhfs}.
The interaction Hamiltonian acts in the Fock space of the electron and nuclear states, but the nuclear states are restricted to the ground-state subspace $\vert I M_I \rangle$ only with $ M_I = -I, \cdots,  I $.
To separate the contributions to the hfs, we restrict ourselves to the effects linear in $H_{\mu}$. In other words, we consider only the Feynman diagrams, where the hyperfine interaction is taken to the first order.
\\
\indent
The hfs parameter $X_a$ according to Eq.~\eqref{hfs3} reads as follows:
\begin{equation}
\label{eq:G}
 X_a  = \dfrac{E(F^+) - E(F^-)}{E_{\rm F}}\,,
\end{equation}
while the energy of an isolated level $E(F)$ of the Li-like ion with total angular momentum $F$ can be found by employing the two-time Green's function method \cite{shabaev:2002:119}:
\be
\label{eq:dE}
 E(F) = \frac{\displaystyle\intG d\varepsilon\,\varepsilon\,G_{F}(\varepsilon)}
    {\displaystyle\intG d\varepsilon\,G_{F}(\varepsilon)}\,.
\ee
The contour $\Gamma$ surrounds only the pole $\varepsilon = E^{(0)}$, where $E^{(0)}$ is the unperturbed energy, which is the sum of the one-electron Dirac and nuclear energies, $G_{F}(\varepsilon) = \langle FM_{F}Ij_a  \vert G(\varepsilon) \vert FM_{F}Ij_a \rangle$, $G(\varepsilon)$ is the two-time Green's function, and $\vert FM_{F}Ij_a \rangle$ is the wave function of the coupled system (nucleus+electrons):
\begin{equation}
\label{eq:w_f}
\vert FM_{F}Ij_a \rangle = \sum_{M_{I}m_a} C^{F M_{F}}_{I M_{I}\ j_am_a}
                           \vert IM_{I} \rangle \vert j_am_a  \rangle\,,
\end{equation}
where $\vert IM_{I} \rangle $ is the nuclear wave function with nuclear spin $I$ and its projection $ M_{I}$, $\vert j_am_a  \rangle$ denotes the unperturbed 3-electron one-determinant wave function in the $1s^22s$ state with the total angular momentum $j_a$ and its projection $m_a$.
\\
\indent
Within the perturbation theory, the energy $E(F)$ and the Green's function $G_F(\varepsilon)$ are to be expanded in the power series in $\alpha$:
\be
\label{eq:E_in_alpha}
  E(F) &=& E^{(0)} +  E^{(1)}(F)  + \dots + E^{(i)}(F) + \dots \,,
\\
\label{eq:G_in_alpha}
 G_{F}(\varepsilon) &=& G^{(0)}(\varepsilon) + G_{F}^{(1)}(\varepsilon) + G_{F}^{(2)}(\varepsilon)  + \dots + G_{F}^{(i)}(\varepsilon)  + \dots \,.
\ee
Here one should note that, in the zeroth order in $\alpha$, there is no interaction of bound electrons with the magnetic field of the nucleus and, therefore, the zeroth-order energy $E^{(0)}$ does not depend on the total angular momentum $F$.
Along this line, the hfs parameter $X_a$ can be expanded in the following way
\be
\nonumber
X_{a} &=& X_{a}^{(0)} + X_{a}^{(1)} + X_{a}^{(2)} + X_{a}^{(3+)}\,,
\\
X_{a}^{(3+)} &=& X_{a}^{(3)} + \dots + X_{a}^{(i)}+ \dots ,
\label{eq:G_hfs}
\ee
where the index $i$ refers to the $i$th order correction $X^{(i)}_a$ in $\alpha$, which can be found as:
\be
\label{eq:K_i}
X_a^{(i)} = \dfrac{E^{(i+1)}(F^+) - E^{(i+1)}(F^-)}{E_{\rm F}}\,.
\ee
Each order in $\alpha$ contains all the relevant corrections, such as the one-electron QED, screened QED, as well as the interelectronic-interaction terms to the hfs. In the present study, however, we restrict ourselves only to the interelectronic-interaction corrections. Then the terms in Eq.~\eqref{eq:G_hfs} refer to the interelectronic-interaction corrections due to the one-photon exchange ($X_{a}^{(1)}$), the two-photon exchange ($X_{a}^{(2)}$) and the higher-order  diagrams ($X_{a}^{(3+)}$), respectively.
 It is worth mentioning, that in contrast to the previous works (see, for example, \cite{volotka:2008:062507, shabaev:1998:149}), we separate out explicitly the two-photon-exchange term $X^{(2)}_a$, since now it is evaluated within the rigorous QED approach. Thus, we represent the interelectronic-interaction correction as a sum of 3 terms, namely, the  one-photon exchange $X^{(1)}_a$, the two-photon exchange $X^{(2)}_a$ and the higher-order term $X^{(3+)}_a$. The first-order term $X^{(1)}_a$ corresponds to the $B(\alpha Z)/Z$ in Refs. \cite{volotka:2008:062507, shabaev:1998:149}, the sum $X^{(2+)}_a = X^{(2)}_a + X^{(3+)}_a$ corresponds to the term $C(\alpha Z,Z)/Z^{2}$ in Refs. \cite{volotka:2008:062507, shabaev:1998:149}.

%
%-----------------------------------------------------------------------
%
\subsection{Zeroth-order contribution}
\label{subsec-A}
The lowest order term $X_{a}^{(0)}$ is given due to Eq.~\eqref{eq:K_i} by:
\be
\label{eq:K_0}
X_a^{(0)} = \dfrac{E^{(1)}(F^+) - E^{(1)}(F^-)}{E_{\rm F}}\,,
\ee
where
\be
\label{eq:dE-1}
   E^{(1)}(F) = \frac{1}{2\pi i}\intG d\varepsilon\,(\varepsilon-E^{(0)})\,G^{(1)}_{F}(\varepsilon).\,
\ee
and $G_{F}^{(1)}(\varepsilon)$ is obtained by using the Feynman rules
\begin{equation}
\label{eq:G_1}
 G_{F}^{(1)}(\varepsilon)
  = \dfrac{\langle FM_{F}Ij_a \vert H_\mu \vert FM_{F}Ij_a \rangle}
          {(\varepsilon - E^{(0)})^2}\,.
\end{equation}
Making use of the Eqs.~\eqref{eq:dE-1} and \eqref{eq:G_1}, we obtain: 
\be
\label{eq:K_0_matel}
 E^{(1)}(F) = \langle FM_{F}Ij_a \vert H_\mu \vert FM_{F}Ij_a \rangle .
\ee
Substituting now Eqs.~\eqref{eq:Hhfs}, \eqref{eq:w_f}, and \eqref{eq:K_0_matel} into Eq.~\eqref{eq:K_0} one can find an explicit form for $X_a^{(0)}$:
\begin{equation}
 X_{a}^{(0)}= G_a \langle a \vert T_0 \vert a \rangle\,,
\end{equation}
which coincides with Eq.~\eqref{K_0} obtained in the one-electron approximation.
\\
\indent
In the present work, we perform calculations within the extended Furry picture, which has been already successfully applied to the QED calculations of various atomic properties \cite{sapirstein:2001:032506, oreshkina:2007:889, sapirstein:2006:042513, oreshkina:2008:675, glazov:2019:062503, sapirstein:2008:022515, sapirstein:2003:022512, ginges:2017:062502, glazov:2006:330, kozhedub:2007:012511, malyshev:2014:062517,kozhedub:2019:062506}. In contrast to the original Furry picture, where the zeroth order Hamiltonian contains only the nuclear potential, in the framework of the extended Furry picture an additional effective local screening potential is added to the unperturbed Hamiltonian.
Hence, the electron wave functions are the eigenstates of the following Dirac Hamiltonian $h^{\rm D}$:
\begin{equation}
\label{eq:dirac_ham}
h^{\rm D}(\br) = \balpha \cdot {\bf p} + \beta + V_{\rm eff}(\br)
\end{equation}
with the effective potential $V_{\rm eff}(\br)$:
\begin{equation}
V_{\rm eff}(\br) = V_{\rm nucl}(\br) + V_{\rm scr}(\br)\,,
\end{equation}
where $V_{\rm nucl}(\br)$ is the electrostatic potential of the nucleus and $V_{\rm scr}(\br)$ is the effective local screening potential. Such an approach allows to accelerate the convergence of the perturbation expansion by accounting part of the interelectronic interaction already in the zeroth order. Moreover, in this way we relieve the quasidegeneracy of the $1s^2 2s$ and $1s^2 2p_{1/2}$ states already at the zeroth-order level and improve the energy level scheme of the first excited states.
In the present work, we employ 3 starting potentials: Coulomb, Core-Hartree, and Kohn-Sham. In order to avoid a double-counting of the screening effects, the counterterm $-V_{\rm scr}(\br)$ should be added to the interaction Hamiltonian, thereby the additional counterterm diagrams appear.
Here, we should note that in previous calculations~\cite{volotka:2008:062507,shabaev:1998:149} the original Furry picture was used, and therefore, only the total value of $X_{a}$ can be compared with corresponding value from Refs.~\cite{volotka:2008:062507,shabaev:1998:149}.
%
%-----------------------------------------------------------------------
%
\subsection{First-order contribution}
\label{subsec-B}
\begin{figure}
\begin{center}
\includegraphics[width=0.25\linewidth]{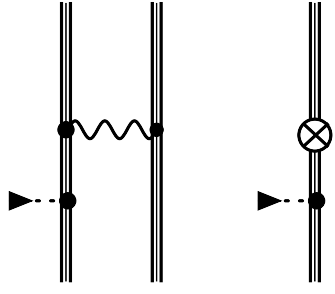}
\caption{Feynman diagrams representing the one-photon exchange correction to the hfs in the framework of the extended Furry picture. The wavy line indicates the photon propagator. The triple line represents the electron propagator in the effective potential $V_{\rm eff}$. The dashed line terminated with the triangle corresponds to the hyperfine interaction. The symbol $\otimes$ represents the extra interaction term associated with the screening potential counterterm.}
\label{ris:one_ph_ex}
\end{center}
\end{figure}

The leading correction $X_{a}^{(1)}$ to the hfs is given by the expression:
\be
X_a^{(1)} = \dfrac{E^{(2)}(F^+) - E^{(2)}(F^-)}{E_{\rm F}}\,,
\ee
where
\be
\label{eq:dE-2}
 E^{(2)}(F) &=& \frac{1}{2\pi i}\intG d\varepsilon\,(\varepsilon-E^{(0)})\,G^{(2)}_{F}(\varepsilon)\nonumber\\
            &-& \frac{1}{2\pi i}\intG d\varepsilon\,(\varepsilon-E^{(0)})\,G^{(1)}_{F}(\varepsilon)\,
                \frac{1}{2\pi i}\intG d\varepsilon\,G^{(1)}_{F}(\varepsilon)\,.
\ee
The one-photon exchange diagrams in the presence of magnetic field of the nucleus corresponding to $G^{(2)}_{F}(\varepsilon)$ are depicted in Fig.~\ref{ris:one_ph_ex}. Within the extended Furry picture, the counterterm diagrams associated with an extra interaction term appear. These diagrams are also shown in Fig.~\ref{ris:one_ph_ex}, where the symbol $\otimes$ denotes local screening potential counterterm.
\\
\indent
Further, employing the Feynman rules for the Green functions $G^{(2)}_{F}(\varepsilon)$ and $G^{(1)}_{F}(\varepsilon)$ and keeping only the linear dependence on $H_\mu$, one obtains the one-photon exchange correction to the hfs in Li-like ion $X_a^{(1)}$ in the form:
\be
\label{B(aZ)}
 X_a^{(1)} &=& 2\,G_a \sum_b  \Bigg[  \sum_P (-1)^P \Biggl( \la \zeta_{b\vert PaPb} | T_0 | a \ra + \la \zeta_{a\vert PbPa} | T_0 | b \ra \Bigg) 
 \nonumber\\ 
 &-& \dfrac{1}{2}\Bigg( \la a | T_0 | a \ra - \la b | T_0 | b \ra \Bigg) \la a b | I^{\prime}(\veps_a-\veps_b) | b a \ra \Bigg] - 2\,G_a\la \eta_a  | T_0 | a \ra
\ee
with
\be
\label{eq:eta}
| \eta_a \ra = {\sum_n}' \frac{| n \ra \la n | V_{\rm scr} | a \ra}{\veps_a-\veps_n}\,,
\ee
\be
\label{eq:zeta}
| \zeta_{a\vert PbPa} \ra = {\sum_n}'
\frac{| n \ra \la n a | I(\Delta) | Pb Pa \ra}{\veps_b-\veps_n}\,,\;\;\; | \zeta_{b\vert PaPb} \ra = {\sum_n}' \frac{| n \ra \la n b | I(\Delta) | Pa Pb \ra}{\veps_a-\veps_n}\,.
\ee
Here $P$ is the permutation operator giving rise to the sign $(-1)^P$ according to the parity of the permutation, $\Delta = \veps_a-\veps_{Pa}$, $\veps_n$ are the one-electron energies, $\vert b \rangle$ stands for the $1s$ state, while the summation over $b$ runs over two possible projections $m_b = \pm 1/2$ of the total angular momentum $j_b$. The prime on the sums over the intermediate states $n$ denotes that the terms with vanishing denominators are omitted. The interelectronic-interaction operator $I(\omega)$ is given by
\begin{equation}
I(\omega, r_{12} ) = \alpha \dfrac{(1-\balpha_{1}\balpha_{2})}{r_{12}}e^{i\tomega r_{12}}
\label{eq:ii_feyn}  
\end{equation}
in the Feynman gauge and by
\begin{equation}
I(\omega, r_{12}) = \alpha \left(  
\dfrac{1}{r_{12}} 
-\dfrac{\balpha_{1}\balpha_{2}}{r_{12}} e^{i\tomega r_{12}}
-\left[ h^{\rm D}_{1},\left[h^{\rm D}_{2},\dfrac{e^{i\tomega r_{12} }-1}
{\omega^{2}r_{12}}\right] \right]\right)
\label{eq:ii_coul}  
\end{equation}
in the Coulomb gauge, $I^{\prime}(\omega) = dI(\omega)/d\omega$, $r_{12}= \vert \bfr_1-\bfr_2 \vert$, and $\tomega=\sqrt{\omega^2+i0}$. The branch of the square root is fixed by the condition ${\rm Im}\,\tomega>0$. We here note that rigorous evaluation of the one-photon exchange correction to the ground-state hfs in Li-like ions was previously performed in Refs.~\cite{volotka:2008:062507, shabaev:1998:149, oreshkina:2007:889} in the framework of the original Furry picture and in Ref.~\cite{sapirstein:2001:032506} in the framework of the extended Furry picture.
%
%-----------------------------------------------------------------------
%
\subsection{Second-order contribution}
\label{subsec-C}
\begin{figure}
\begin{center}
\includegraphics[width= 0.7\linewidth]{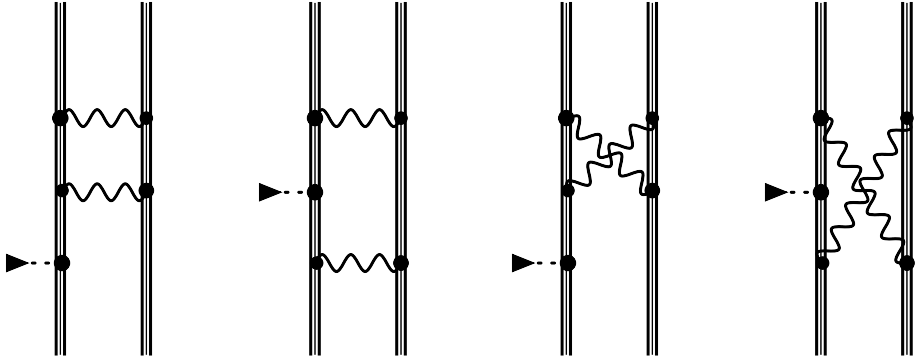}
\caption{Feynman diagrams representing the two-electron part of the two-photon-exchange correction to the hfs in the framework of the extended Furry picture. Notations are the same as in Fig.~\ref{ris:one_ph_ex}.}
\label{fig-2el}
\end{center}
\end{figure}

\begin{figure}
\begin{center}
\includegraphics[width= 0.7\linewidth]{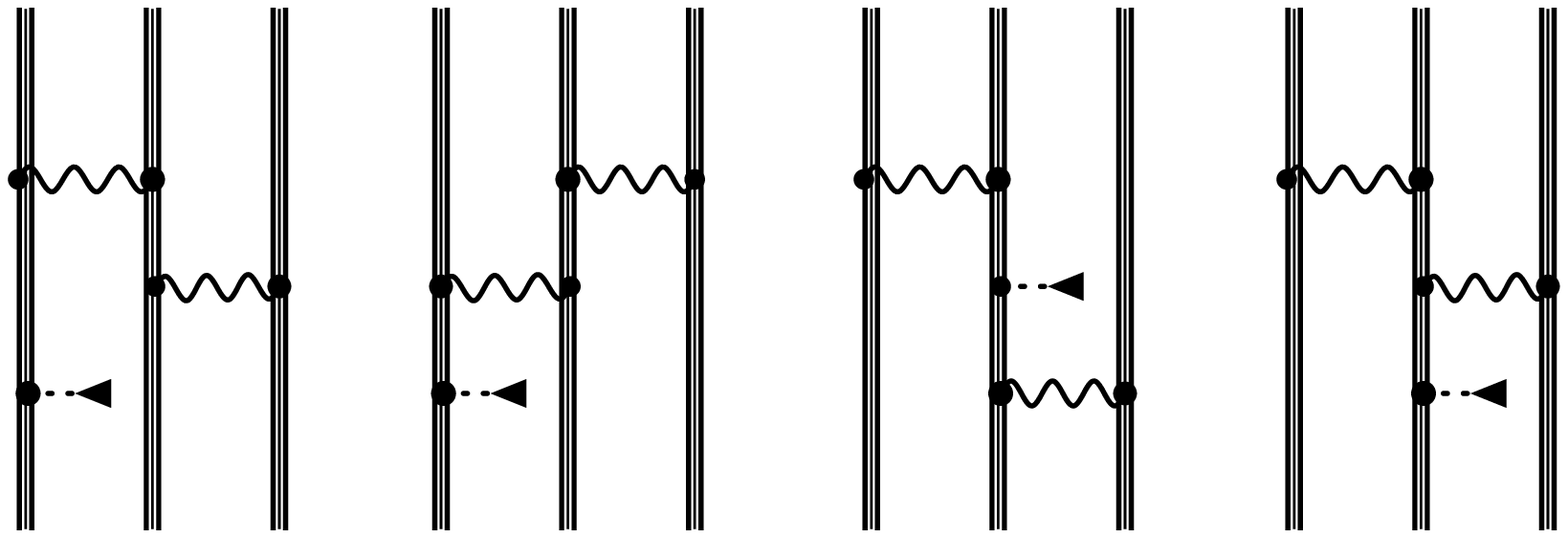}
\caption{Feynman diagrams representing the three-electron part of the two-photon-exchange correction to the hfs in the framework of the extended Furry picture. Notations are the same as in Fig.~\ref{ris:one_ph_ex}.}
\label{fig-3el}
\end{center}
\end{figure}

\begin{figure}
\begin{center}
\includegraphics[width= 0.8\linewidth]{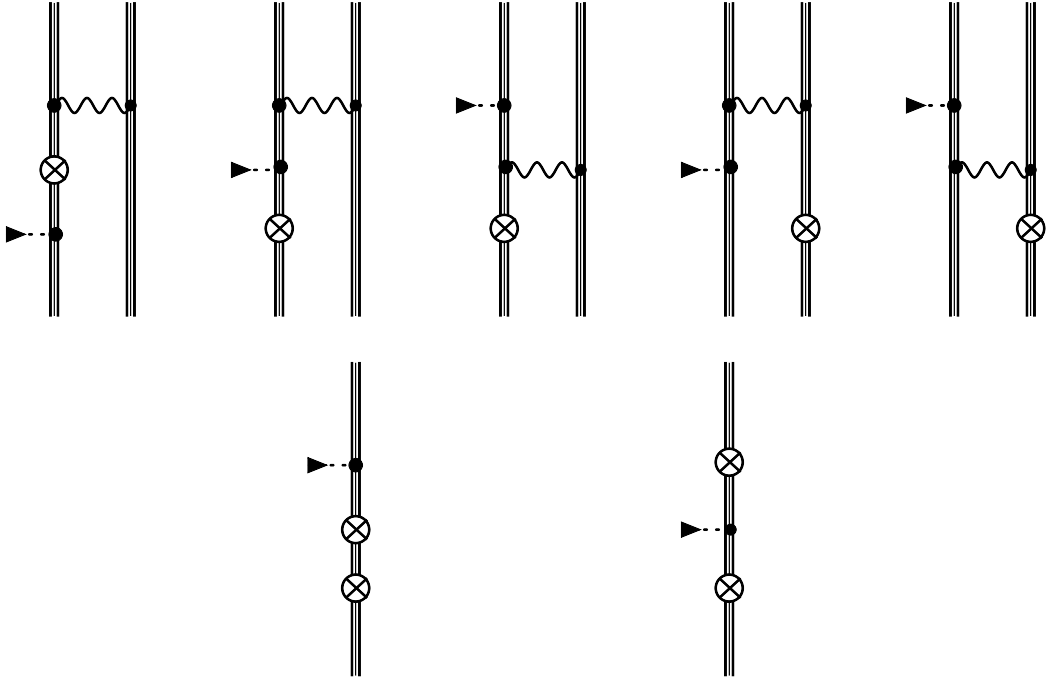}
\caption{Feynman diagrams representing the counterterm part of the two-photon-exchange correction to the hfs in the framework of the extended Furry picture. Notations are the same as in  Fig.~\ref{ris:one_ph_ex}.}
\label{fig-countertems}
\end{center}
\end{figure}

 The second-order correction $X_{a}^{(2)}$ to the hfs can be found as
\be
X_a^{(2)} = \dfrac{E^{(3)}(F^+) - E^{(3)}(F^-)}{E_{\rm F}}\,,
\ee
where
\be
\label{eq:dE-3}
 E^{(3)}(F) &=& \frac{1}{2\pi i}\intG d\varepsilon\,(\varepsilon-E^{(0)})\,G^{(3)}_{F}(\varepsilon)
 - \frac{1}{2\pi i}\intG d\varepsilon\,(\varepsilon-E^{(0)})\,G^{(2)}_{F}(\varepsilon)\,
   \frac{1}{2\pi i}\intG d\varepsilon\,G^{(1)}_{F}(\varepsilon) \nonumber\\
&-&\frac{1}{2\pi i}\intG d\varepsilon\,(\varepsilon-E^{(0)})\,G^{(1)}_{F}(\varepsilon)
    \left[\frac{1}{2\pi i}\intG d\varepsilon\,G^{(2)}_{F}(\varepsilon)
         -\left(\frac{1}{2\pi i}\intG d\varepsilon\,G^{(1)}_{F}(\varepsilon)\right)^2
    \right]\,.
\ee
The diagrams of the two-photon exchange in the presence of magnetic field of the nucleus corresponding to the Green's function $G^{(3)}_{F}(\varepsilon)$ are given in Figs.~\ref{fig-2el}, \ref{fig-3el}, and \ref{fig-countertems}.
These diagrams are divided into three groups: the two-electron (Fig.~\ref{fig-2el}), the three-electron (Fig.~\ref{fig-3el}), and the counterterm (Fig.~\ref{fig-countertems}) ones.
The formal expressions for the first two groups were derived in Ref.~\cite{volotka:2012:073001} and, therefore, we do not present them here. However, in contrast to the original Furry picture employed in Ref.~~\cite{volotka:2012:073001}, in the extended Furry picture additional counterterm diagrams have to be taken into account. The formal expressions corresponding to the counterterm diagrams depicted in Fig.~\ref{fig-countertems} can be found in Appendix A.
\\ 
\indent
The two-electron contributions involve the integration over the energy of the virtual photon $\omega$, which for some terms is infrared-divergent. These infrared-divergent terms can be, however, separated and the divergences cancel each other out. In order to avoid strong oscillations arising for large real values of $\omega$, we perform a Wick rotation with the integration contours chosen as in Ref. \cite{mohr:2000:052501}.
Moreover, the formal expressions for the two-photon exchange correction involve infinite summations over the complete Dirac spectrum including the infinite partial wave expansion. The summation over the intermediate states is performed by employing the dual-kinetic-balance finite basis set method \cite{shabaev:2004:130405} for the Dirac equation. The basis functions are constructed from B-splines \cite{sapirstein:1996:5213}. We systematically increase the number of basis functions from $N = 92$ to $N = 212$ to achieve a clear convergence pattern of the calculated results and then performed the extrapolation $N\rightarrow\infty$ .
The partial wave summation over the Dirac quantum number $\kappa$ was terminated at $\kappa_{\rm max} = 10$ and the remainder was estimated using least-squares inverse polynomial fitting. The absolute uncertainty of this estimation in $X_{a}^{(2)}$ is found to be around $3 \times 10^{-7}$ in the case of $Z = 7$ and rapidly decrease to $10^{-7}$ or smaller with the increase of $Z$.
\\ 
\indent
For a consistency check, the two-photon exchange correction is calculated within the Feynman and Coulomb gauges, and the results are found to be gauge invariant to a very high accuracy. As an another consistency check, we compare the obtained results for the two-photon exchange correction with the results evaluated within the Breit approximation (see the next subsection). The comparison comprises both the numerical check as well as analytical, which was performed by replacing the interelectronic-interaction operator $I(\omega)$ to its frequency-independent counterpart in the Breit approximation.
All of this confirms the reliability of the present calculations. 
%
%-----------------------------------------------------------------------
%
\subsection{Third- and higher-orders contribution}
\label{subsec-D}
While the rigorous QED approach is bound at the moment to the first and second orders of the interelectronic interaction, the third- and higher-order contributions ($X_{a}^{(3+)}$) are also important at the present level of theoretical accuracy. So, these contributions are presently considered within the so-called Breit approximation based on the Dirac-Coulomb-Breit equation. Within this approximation, the interelectronic-interaction operator $I(\omega)$ is replaced by its $\omega=0$ -- limit taken in the Coulomb gauge:
\begin{equation}
  I(\omega, r_{12}) \rightarrow I_{\rm B}(r_{12}) = \alpha \left( \dfrac{1}{r_{12}} 
    - \frac{\balpha_{1}\balpha_{2}}{r_{12}} 
    + \frac{1}{2}\left[h^{\rm D}_{1},\left[h^{\rm D}_{2},r_{12}\right] \right]
  \right)\,.
\label{eq:breit0} 
\end{equation}
Another important component of the Dirac-Coulomb-Breit approach is the projector on the positive-energy states with respect to the one-electron Dirac Hamiltonian. Its origin was demonstrated in Refs.~\cite{faustov:1970:478, sucher:1980:348}. The use of this projector leads effectively to the suppression of the processes involving the virtual electron-positron pairs, therefore it is also referred to as the ``no-pair approximation'' (Breit$_{\rm no-ee^+}$). However, the mixing of the large and small components of the Dirac wave functions by the Fermi-Breit operator~\eqref{eq:Hhfs}, \eqref{bfT} (and any other operator with $\balpha$-matrices) leads to the strong enhancement of the negative-energy contributions. In order to take these contributions into account, one should either construct the positive-energy projectors with respect to the Dirac Hamiltonian which includes the Fermi-Breit operator or evaluate these contributions separately as the first-order perturbation in this operator. We use the second approach, which corresponds to the inclusion of the processes involving one virtual electron-positron pair (Breit$_{\rm one-ee^+}$).
\\ 
\indent
The interelectronic interaction can be taken into account within any of the available methods \cite{boucard:2000:59, dzuba:1987:1399, ginges:2018:032504, shabaev:1995:3686, zherebtsov:2000:701, yerokhin:2008:R020501, blundell:1989:2233, bratsev:1977:2655, glazov:2017:46}. Previously, the all-order CI-DFS method \cite{bratsev:1977:2655} was employed to evaluate $X_{a}^{(3+)}$ \cite{volotka:2012:073001} or $X_{a}^{(2+)}$~\cite{volotka:2008:062507}. In this work, we opt for the recursive formulation of the perturbation theory \cite{glazov:2017:46}. This method allows one to access efficiently the individual terms of the perturbation expansion up to any order. It also ensures that the zeroth-order Hamiltonian is the same for the rigorous QED and Breit-approximation calculations. Recently, this method has been successfully applied to the similar calculations of the higher-order contributions to the $g$ factor of Li-like ions \cite{glazov:2019:173001}.

To formulate this approach, we start with the Dirac-Coulomb-Breit equation,
\begin{equation}
\label{eq:DCB}
  \Lambda_+ \left( H_0 + H_1 \right) \Lambda_+ \ket{A} = E_A \ket{A}
\,,
\end{equation}
where $\Lambda_+$ is the positive-energy-states projection operator, constructed as the product of the one-electron projectors. The zeroth-order Hamiltonian $H_0$ is the sum of the one-electron Dirac Hamiltonians,
\begin{equation}
\label{eq:H_0}
  H_0 = \sum_j h^{\rm D}(\br_j)
\,,
\end{equation}
where $h^{\rm D}$ is given by Eq.~(\ref{eq:dirac_ham}). The eigenfunctions of $H_0$,
\begin{equation}
\label{eq:A_0}
  \Lambda_+ H_0 \Lambda_+ \ket{N^{(0)}} = E_N^{(0)} \ket{N^{(0)}}
\,,
\end{equation}
form the orthogonal basis set of the many-electron wave functions. $\ket{N^{(0)}}$ can be constructed as the Slater determinants of the one-electron solutions of the Dirac equation. For the reference state $\ket{A}$ in the zeroth approximation we have
\begin{equation}
\label{eq:A_0}
  \Lambda_+ H_0 \Lambda_+ \ket{A^{(0)}} = E_A^{(0)} \ket{A^{(0)}}
\,.
\end{equation}
$H_1$ in Eq.~(\ref{eq:DCB}) represents the interelectronic interaction in the Breit approximation with the screening potential subtracted,
\begin{equation}
  H_1 = \,\sum_{j<k} I_{\rm B}(r_{jk}) - \sum_j V_{\rm scr}(\br_j)
\,.
\end{equation}
We use the perturbation theory with respect to $H_1$, which yields the following expansions for the energy $E_A$ and the wave function $\ket{A}$,
\begin{align}
\label{eq:ten}
  E_A &= \sum_{k=0}^{\infty} E_A^{(k)}
\,,\\
\label{eq:ta}
  \ket{A}
   &= \sum_{k=0}^{\infty} \ket{A^{(k)}}
    = \sum_{k=0}^{\infty} \sum_N \ket{N^{(0)}} \braket{N^{(0)}}{A^{(k)}}
\,.
\end{align}
In Ref.~\cite{glazov:2017:46} the recursive scheme to evaluate $E_A^{(k)}$ and $\braket{N^{(0)}}{A^{(k)}}$ order by order was presented. Here, we consider how to find the contributions $X_{a}^{(k)}$. Substituting Eq.~(\ref{eq:ta}) into the obvious relation
\begin{equation}
\label{eq:XA}
  X_{a} = G_a \, \matrixel{A}{T_0}{A}
% \,,
\end{equation}
we find
\begin{align}
\label{eq:XAk}
  X_{a}^{(k)} &= G_a \, \sum_{j=0}^{k} \matrixel{A^{(j)}}{T_0}{A^{(k-j)}}
\nonumber\\
  &= G_a \, \sum_{j=0}^{k} \sum_{M,N} \braket{A^{(j)}}{M^{(0)}} \matrixel{M^{(0)}}{T_0}{N^{(0)}} \braket{N^{(0)}}{A^{(k-j)}}
\,.
\end{align}
We note that the normalization condition $\braket{A}{A}=1$ is used here instead of the widely accepted intermediate normalization $\braket{A^{(0)}}{A}=1$. Eq.~(\ref{eq:XAk}) is used to find $X_{a}^{(k)}$ in the no-pair Breit approximation. Within the one-pair Breit approximation we add the contribution of the negative-energy excitations, which is found as
\begin{equation}
\label{eq:XAneg}
  X_{a}[-] = 2 G_a \, \sum_{p,n} \frac{\matrixel{p}{T_0}{n}}{\veps_p-\veps_n}
             \matrixel{\hat a^+_n \hat a_p A}{H_1}{A}
\,.
\end{equation}
Here $\ket{p}$ and $\ket{n}$ are the positive- and negative-energy one-electron states, respectively, $\hat a^+$ and $\hat a$ are the corresponding creation and annihilation operators. The corresponding contribution of the order $k$ is
\begin{equation}
\label{eq:XAnegk}
  X_{a}^{(k)}[-] = 2 G_a \, \sum_{j=0}^{k-1} \sum_{M,N} \braket{A^{(j)}}{M^{(0)}}
    \left[ \sum_{p,n} \frac{\matrixel{p}{T_0}{n} \matrixel{\hat a^+_n \hat a_p M^{(0)}}{H_1}{N^{(0)}}}{\veps_p-\veps_n} \right] 
  \braket{N^{(0)}}{A^{(k-j-1)}}
\,.
\end{equation}
We use Eqs.~(\ref{eq:XAk}) and (\ref{eq:XAnegk}) to find the required third- and higher-order contributions from the wave-function coefficients $\braket{N^{(0)}}{A^{(k)}}$ obtained within the recursive scheme.
% The energy corrections $E_A^{(k)}$ and the coefficients $\braket{N^{(0)}}{A^{(k)}}$ can be found via the recursive system of equations,

%
% ==================== RESULTS AND DISCUSSIONS =========================
%
\section{RESULTS AND DISCUSSIONS}
\label{sec-3}
\begin{figure}
\begin{center}
\includegraphics[width=1\linewidth]{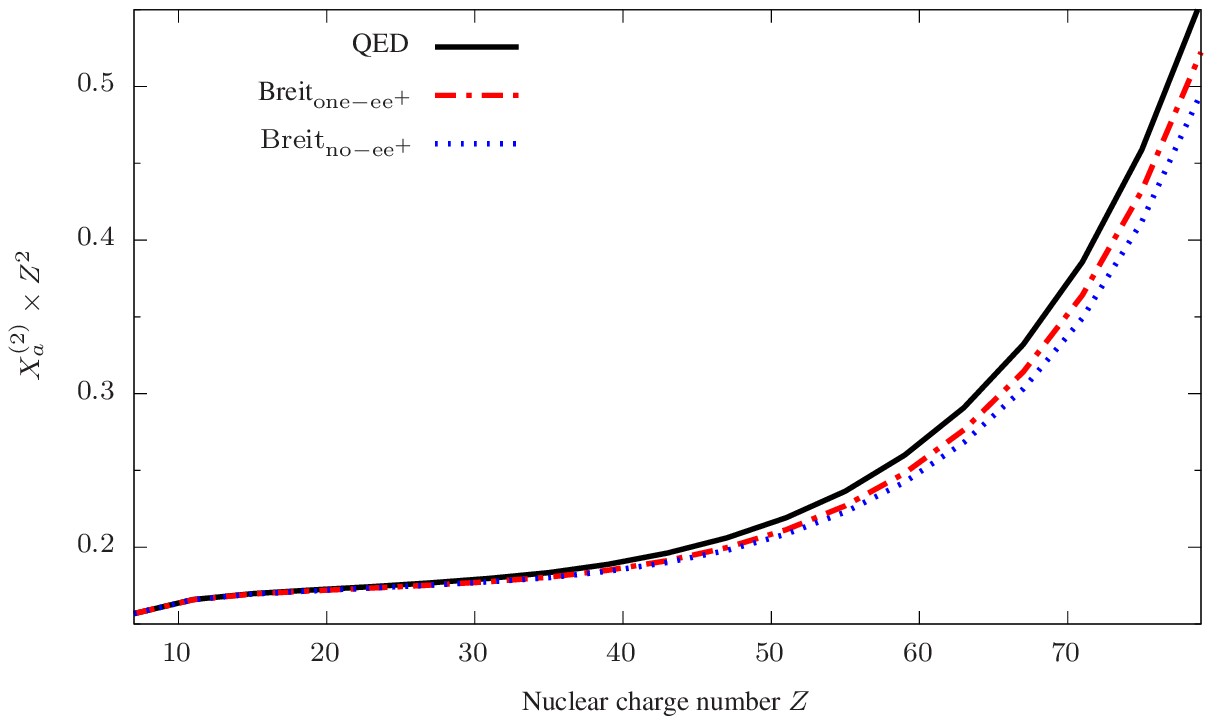}
\caption{The two-photon-exchange correction $X_{a}^{(2)}$ to the ground-state hfs in Li-like ions scaled by a factor $Z^2$ as a function of the nuclear charge number. The results of the rigorous QED calculation (solid black) are compared within the Breit approximation calculations: no-pair (Breit$_{\rm no-ee^+}$, red dashed) and one-pair (Breit$_{\rm one-ee^+}$, blue dotted). All values are obtained with the Kohn-Sham starting potential.}
\label{ris:G_2_KS}
\end{center}
\end{figure}

At first, we discuss the nuclear models and nuclear parameters employed in the calculations. The finite size of the nucleus is accounted for within the Fermi model for the nuclear charge density with the charge radii taken from Ref.~\cite{angeli:2013:69}. The Bohr-Weisskopf correction $\epsilon$ is calculated within the homogeneous sphere model assuming the same radius as used for the charge distribution. Here, we note, that the ratio of the Bohr-Weisskopf corrections calculated with different magnetization distribution models stays the same to a good accuracy for Li-like ions \cite{shabaev:2001:3959} or even for neutral atoms \cite{ginges:2018:032504} as obtained for H-like ions. With this in mind, we can use the results for H-like ions, for example, obtained with the odd nucleon model \cite{shabaev:1997:252} $\epsilon_{\rm odd}^{\rm H}$, to evaluate the Bohr-Weisskopf correction for the corresponding Li-like ion, i.e., $\epsilon_{\rm odd}^{\rm Li} = \epsilon_{\rm odd}^{\rm H}(\epsilon_{\rm sph}^{\rm Li}/\epsilon_{\rm sph}^{\rm H})$.
\\ 
\indent
In Tables~\ref{table:var_pot1}, \ref{table:var_pot2}, and \ref{table:var_pot3} the interelectronic-interaction corrections to the ground state hfs in Li-like $^{15}$N$^{4+}$, $^{98}$Tc$^{40+}$, and $^{209}$Pb$^{79+}$ ions, respectively, are presented including individual orders of the perturbation theory. The uncertainties of the individual terms are determined by the convergence with respect to the numbers of the basis functions and its maximum orbital momentum. The results are obtained with three starting potentials: Coulomb, Core-Hartree, and Kohn-Sham. The latter two correspond to the extended Furry picture and allow to take partially into account the interelectronic interaction already in zeroth order.
 In Tables~I and III, we also compare our results with the corresponding terms from the previous theoretical calculations~\cite{volotka:2008:062507} and~\cite{shabaev:1998:149}, respectively, for the case of the original Furry picture. Namely, the one-electron relativistic factor $X_{a}^{(0)}$ refers to the product $A(\alpha Z)(1-\delta)$ from Refs.~\cite{volotka:2008:062507,shabaev:1998:149}, one-photon exchange correction $X_{a}^{(1)}$ and the higher-order terms $X_{a}^{(2+)}$, correspond to the notations $B(\alpha Z)/Z$ and $C(\alpha Z, Z)/Z^{2}$ or $C(0)/Z^{2}$ from Refs.~\cite{volotka:2008:062507,shabaev:1998:149}, respectively. 
Here we want to stress that we have corrected the values from Ref.~\cite{volotka:2008:062507,shabaev:1998:149} (see third column in Tables~I,~III) to the Fermi model of the charge distribution with the radii taken from Ref.~\cite{angeli:2013:69} and to the point-like magnetization distribution as it is calculated in the present work. Therefore we can conclude that the main reason for the deviation of our values from the ones calculated in~\cite{volotka:2008:062507,shabaev:1998:149} is due to the different treatment of the two-photon-exchange $[X_{a}^{(2)}]$ and the higher-order $[X_{a}^{(3+)}]$ terms.
Indeed, from the Tables~I, III one can see that $X_{a}^{(0)}$ and $X_{a}^{(1)}$ are in a good agreement with the ones, obtained in Refs.~\cite{volotka:2008:062507,shabaev:1998:149}, while the higher orders $X_{a}^{(2+)}$ are improved in comparison to Ref.~\cite{volotka:2008:062507} mainly due to the recursive perturbation theory employed in the present investigation and in comparison to the Ref.~\cite{shabaev:1998:149} due to the rigorous evaluation of the two-photon-exchange correction.
The framework of the extended Furry picture enhances the convergence in comparison to the perturbation theory based on the Coulomb starting potential (original Furry picture). As one can see from Tables~\ref{table:var_pot1}, \ref{table:var_pot2}, and \ref{table:var_pot3} the employment of the extended Furry picture improves the accuracy of interelectronic-interaction correction especially in the low-$Z$ region. For example, in the case of nitrogen $^{15}$N$^{4+}$ the uncertainty of the total value is improved by a factor of four.
\begin{table}
\caption{Interelectronic-interaction contributions to the ground-state hfs in Li-like $^{15}$N$^{4+}$ with various starting potentials: Coulomb, Core-Hartree, and Kohn-Sham, in terms of $X_{a}$, defined by Eq.~\eqref{eq:K_i}.}
\label{table:var_pot1}
\centering
\tabcolsep6pt
%\begin{tabular}{c S S S S S}
\begin{tabular}{clllll}
\hline\hline
 &\multicolumn{3}{c}{Coulomb} &\multicolumn{1}{c}{Core-Hartree} &\multicolumn{1}{c}{Kohn-Sham}\\
  &\multicolumn{1}{c}{This work} &\multicolumn{1}{c}{Ref.~\cite{volotka:2008:062507}$^{ a}$ } &\multicolumn{1}{c}{Ref.~\cite{volotka:2008:062507}} &\multicolumn{1}{c}{This work} &\multicolumn{1}{c}{This work}\\
\hline
$X_{a}^{(0)}$   &\,\,1.004\,912      & \,\,1.004\,91   & \,\,1.004\,89     & \,\,0.617\,954       & \,\,0.618\,795    \\
$X_{a}^{(1)}$   &-0.381\,459         & -0.381\,46      & -0.381\,01        & \,\,0.023\,225       & \,\,0.019\,646    \\
$X_{a}^{(2)}$   &\,\,0.018\,867      &            &                        & \,\,0.000\,249       & \,\,0.003\,200 \\
$X_{a}^{(3)}$   &-0.001\,027(12)     &            &                        & \,\,0.000\,085(3)    &-0.000\,172(3) \\
$X_{a}^{(4)}$   &\,\,0.000\,139(8)   &            &                        &-0.000\,022(7)        & \,\,0.000\,033(4) \\
$X_{a}^{(5)}$   &\,\,0.000\,048(2)   &            &                        & \,\,0.000\,006(3)    &-0.000\,006(1) \\
$X_{a}^{(6)}$   &\,\,0.000\,013(1)   &            &                        &-0.000\,001(1)        & \,\,0.000\,001    \\
$X_{a}^{(7)}$   &\,\,0.000\,003(1)   &            &                        & \,\,0.000\,001       &-0.000\,000    \\
$X_{a}^{(3+)}$  &-0.000\,825(15)     &            &                        & \,\,0.000\,068(8)    &-0.000\,144(5) \\
$X_{a}^{(2+)}$  &\,\,0.018\,042(15)  & \,\,0.018\,00   & \,\,0.017\,98     & \,\,0.000\,317(8)    & \,\,0.003 \,056(5)      \\
Total           &\,\,0.641\,495(15)  & \,\,0.641\,45   & \,\,0.641\,86     & \,\,0.641\,496(8)    &\,\,0.641\,497(5) \\
\hline\hline
\end{tabular}
\\[2mm]
\footnotesize{$^{ a}$ Results from Ref.~\cite{volotka:2008:062507}, recalculated to the nuclear models and nuclear parameters employed in this paper.}\\
\end{table}
%
%\footnotesize{$^{ a}$ Results from Ref.~\cite{volotka:2008:062507}, recalculated to the nuclear charge and magnetic distribution employed in this paper}\\
%
% 
\begin{table}
\caption{Interelectronic-interaction contributions to the ground-state hfs in Li-like $^{98}$Tc$^{40+}$ with various starting potentials: Coulomb, Core-Hartree, and Kohn-Sham, in terms of $X_{a}$, defined by Eq.~\eqref{eq:K_i}.}
\label{table:var_pot2}
\centering
\begin{tabular}{c S S S}
\hline\hline
 &\multicolumn{1}{c}{Coulomb} &\multicolumn{1}{c}{Core-Hartree} &\multicolumn{1}{c}{Kohn-Sham}\\
\hline
$X_{a}^{(0)}$	& 1.233 403 0      & 1.145 748 3     & 1.148 599 7 \\
$X_{a}^{(1)}$	&-0.077 938 0      & 0.010 485 0     & 0.007 506 6 \\
$X_{a}^{(2)}$	& 0.000 755 7      &-0.000 022 3     & 0.000 106 1 \\
$X_{a}^{(3)}$	&-0.000 008 6      & 0.000 001 2     &-0.000 000 5            \\
$X_{a}^{(4)}$	& 0.000 000 1      &-0.000 000 1	 & 0.000 000 0             \\
$X_{a}^{(3+)}$  &-0.000 008 4   & 0.000 001 2  &-0.000 000 5 \\
Total           & 1.156 212 3   & 1.156 212 2  & 1.156 212 0 \\
\hline\hline
\end{tabular}
\end{table} 
% 
%\begin{table}
%\caption{Interelectronic-interaction contributions to the ground-state hfs in Li-like $^{197}$Au$^{76+}$ with various starting potentials: Coulomb, Core-Hartree, and Kohn-Sham, in terms of $X_{a}$, defined by Eq.~\eqref{eq:K_i}.}
%\label{table:var_pot3}
%\centering
%\begin{tabular}{c S S S}
%\hline\hline
% &\multicolumn{1}{c}{Coulomb} &\multicolumn{1}{c}{Core-Hartree} &\multicolumn{1}{c}{Kohn-Sham}\\
%\hline
%$X_{a}^{(0)}$	& 2.227 141 6      & 2.127 038 4      & 2.134 320 5	\\
%$X_{a}^{(1)}$	&-0.082 093 1      & 0.018 680 8      & 0.011 326 4 \\
%$X_{a}^{(2)}$	& 0.000 694 6      & 0.000 015 9      & 0.000 089 0 \\
%$X_{a}^{(3)}$	&-0.000 007 7(1)      & 0.000 000 5      &-0.000 000 3 \\
%$X_{a}^{(4)}$	& 0.000 000 1      & 0.000 000 0      & 0.000 000 0 \\
%$X_{a}^{(3+)}$  &-0.000 007 6(1)   & 0.000 000 5   &-0.000 000 3 \\
%Total		    & 2.145 735 5(1)   & 2.145 735 6   & 2.145 735 7 \\
%\hline\hline
%\end{tabular}
%\end{table} 
%
\begin{table}
\caption{Interelectronic-interaction contributions to the ground-state hfs in Li-like $^{209}$Pb$^{79+}$ with various starting potentials: Coulomb, Core-Hartree, and Kohn-Sham, in terms of $X_{a}$, defined by Eq.~\eqref{eq:K_i}.}
\label{table:var_pot3}
\centering
\tabcolsep6pt
\begin{tabular}{clllll}
\hline\hline
 &\multicolumn{3}{c}{Coulomb} &\multicolumn{1}{c}{Core-Hartree} &\multicolumn{1}{c}{Kohn-Sham}\\
  &\multicolumn{1}{c}{This work} &\multicolumn{1}{c}{Ref.~\cite{shabaev:1998:149}$^{ a}$ } &\multicolumn{1}{c}{Ref.~\cite{shabaev:1998:149}} &\multicolumn{1}{c}{This work} &\multicolumn{1}{c}{This work}\\
\hline
$X_{a}^{(0)}$   &\,\,2.397\,606\,5      &\,\,2.397\,6   &\,\,2.398\,7    &2.292\,003\,7        &\,\,2.300\,174\,6    \\
$X_{a}^{(1)}$   & -0.085\,899\,5        & -0.085\,9     & -0.081\,7      &0.020\,407\,4        &\,\,0.012\,164\,4     \\
$X_{a}^{(2)}$   &\,\,0.000\,736\,3      &               &                &0.000\,023\,5        &\,\,0.000\,096\,6 \\
$X_{a}^{(3)}$   & -0.000\,008\,2        &               &                &0.000\,000\,5        &-0.000\,000\,4             \\
$X_{a}^{(4)}$   &\,\,0.000\,000\,1      &               &                &0.000\,000\,0        &\,\,0.000\,000\,0             \\
$X_{a}^{(3+)}$  & -0.000\,008\,1        &               &                &0.000\,000\,5        &-0.000\,000\,4 \\
$X_{a}^{(2+)}$  &\,\,0.000\,728\,2      &\,\,0.000\,1   &\,\,0.000\,1    &0.000\,024\,0        & 0.000\,096\,2 \\
Total           &\,\,2.312\,435\,2      &\,\,2.311\,8   &\,\,2.317\,1    &2.312\,435\,1        &\,\,2.312\,435\,2\\
\hline\hline
\end{tabular}
\\[2mm]
\footnotesize{$^{ a}$ Results from Ref.~\cite{shabaev:1998:149}, recalculated to the nuclear models and nuclear parameters employed in this paper.}\\
\end{table}
\\ 
\indent
The Breit and QED treatments of the one- [$X_{a}^{(1)}$] and two-photon exchange [$X_{a}^{(2)}$] corrections to the ground-state hfs in Li-like ions are compared in Table~\ref{table:MBPT_vs_QED} and Fig.~\ref{ris:G_2_KS}. The values are obtained within the extended Furry picture with the Kohn-Sham potential. Within the Breit approximation we distinguish two results: "no-pair" (Breit$_{\rm no-ee^+}$) and "one-pair" (Breit$_{\rm one-ee^+}$), see Section \ref{subsec-D}. From the results presented in Table~\ref{table:MBPT_vs_QED} and Fig.~\ref{ris:G_2_KS} it can be seen that for light ions the difference between QED approach and both Breit approximations (Breit$_{\rm no-ee^+}$/Breit$_{\rm one-ee^+}$) is less than 0.1\%, but it increases fast with the growth of $Z$. In particular, for gold ($Z = 79$) the deviation between the rigorous QED treatment and the Breit$_{\rm one-ee^+}$ approximation for the two-photon exchange term is about 6\%, while the QED--Breit$_{\rm no-ee^+}$ difference is more than 12\%.
\begin{table}
\caption{Comparison of the one- [$X_{a}^{(1)}$] and two-photon exchange [$X_{a}^{(2)}$] corrections to the ground-state hfs in Li-like ions calculated within the rigorous QED approach and within the Breit approximations: no-pair (Breit$_{\rm no-ee^+}$) and one-pair (Breit$_{\rm one-ee^+}$), see text for details. The values are obtained with the Kohn-Sham starting potential.}
\label{table:MBPT_vs_QED}
\centering
\begin{tabular}{c c S S S}
\hline\hline
Ion	   &       &\multicolumn{1}{c}{QED}
               &\multicolumn{1}{c}{Breit$_{\rm one-ee^+}$}	
               &\multicolumn{1}{c}{Breit$_{\rm no-ee^+}$}                         \\
\hline
$^{15}$N$^{4+}$    &$X_{a}^{(1)}$ &0.019 646 3(2) &0.019 645 9(2) &0.019 641 2(2) \\
                   &$X_{a}^{(2)}$ &0.003 199 9(3) &0.003 198 2(3) &0.003 198 5(3) \\[1mm]
$^{98}$Tc$^{40+}$  &$X_{a}^{(1)}$ &0.007 506 6    &0.007 489 1    &0.007 435 5    \\
                   &$X_{a}^{(2)}$ &0.000 106 1    &0.000 103 5    &0.000 102 8    \\[1mm]
$^{197}$Au$^{76+}$ &$X_{a}^{(1)}$ &0.011 326 4    &0.011 169 9    &0.011 224 4    \\
                   &$X_{a}^{(2)}$ &0.000 089 0    &0.000 083 7    &0.000 079 4    \\
\hline\hline
\end{tabular}
\end{table} 
\\ 
\indent
In Table~\ref{table:final_results} the interelectronic-interaction contributions calculated with the Kohn-Sham starting potential are presented in terms of the dimensionless hfs parameter $X_a$ for the wide range of the nuclear charge number $Z = 7 - 82$. The zeroth-order values $X_a^{(0)}$ are shown in the second column, while the contributions of the first order ($X_a^{(1)}$), second order ($X_a^{(2)}$), and higher orders ($X_a^{(3+)}$) are listed in columns three to five, respectively. In the last two columns, the total value of $X_a$ and the Bohr-Weisskopf correction evaluated within the homogeneous sphere model are given. The uncertainty of the total value is determined as a root-sum-square of the numerical uncertainties of the individual corrections and the unknown QED contribution in the third order in $1/Z$. 
Strictly  speaking,  the  third-order Breit approximation is valid up to the order $(\alpha Z)^2$, and therefore the  treatment of the $X_a^{(3+)}$ term in the framework of the recursive perturbation theory based on the Breit approximation is also valid up to the order $(\alpha Z)^2$. Thus, in the present work the unknown QED contribution of the third order in $1/Z$ is estimated as $(\alpha Z)^3/Z^3$.
\\ 
\indent
In Table~\ref{table:final_results}, we also compare our total values of the interelectronic-interaction contribution with the corresponding results of the previous theoretical calculations \cite{volotka:2008:062507, shabaev:1998:149}. Here, we want to stress that the values of the total interelectronic-interaction correction obtained in Ref.~\cite{volotka:2008:062507,shabaev:1998:149} are corrected to the nuclear models and nuclear parameters employed in the present work.
Here we notice that in the framework of the extended Furry picture we can not perform comparison with Refs. \cite{volotka:2008:062507, shabaev:1998:149} term by term, in contrast only the total values (see column 6 in Table~V), can be compared.
\\ 
\indent
From Table V one can see that for the light ions, for example, $^{15}$N$^{4+}$, the diviation between present results and the ones given in Ref.~\cite{volotka:2008:062507} is about 0.006\% and it decreases fast with the growth of the nuclear charge number $Z$. It can be explained by the fact that in the framework of the perturbation theory within the extended Furry picture employed in the present work, the higher-order corrections $X_a^{(3+)}$ converge faster with the growth of $Z$. 
%Therefore, already for the case of $^{51}$V$^{20+}$ we observe a generally good agreement of our predictions with those obtained in Ref.~\cite{volotka:2008:062507}, where the second- and higher-order contributions were evaluated in the Breit approximation within the CI-DFS method.
%
We also compare our results with Ref.~\cite{shabaev:1998:149} for the high-$Z$ region. As one can see from Table V, the deviation between the present results and Ref.~\cite{shabaev:1998:149} is about 0.02\% -- 0.03\%, much larger than in the middle-$Z$ region. It is mainly explained by the fact that in the work \cite{shabaev:1998:149} the correction $X_{a}^{(2)}$ was estimated by its nonrelativistic limit and the contribution $X_{a}^{(3+)}$ was not taken into account.
Moreover, in contrast to the previous calculations \cite{volotka:2008:062507, shabaev:1998:149} we have more carefully analyzed all the uncertainties , i.e. numerical error of the individual terms and the unknown higher-order contributions. 
\\ 
\indent
The present results alone do not improve the hfs values of Li-like ions, since the uncertainty of the Bohr-Weisskopf correction dominates through all the nuclear charge range under consideration.  However, this uncertainty can be significantly reduced in the specific differences of the hfs values for different charge states \cite{shabaev:2001:3959,volotka:2014:023002}. Simultaneous evaluation of the screened QED corrections is also in demand to obtain the most accurate theoretical predictions for the specific differences.
\begin{table}
\caption{Interelectronic-interaction contributions to the ground-state hfs in Li-like ions obtained with the Kohn-Sham potential, in terms of the hfs parameter $X_{a}$ defined by Eq.~\eqref{hfs3}. In the last two columns, the total value of $X_a$ and the Bohr-Weisskopf correction evaluated within the homogeneous sphere model are also presented. The uncertainty of the total value (numbers in parentheses) is determined as a root-sum-square of the numerical uncertainties of the individual corrections and the unknown QED contribution of the third order in $1/Z$ estimated as $(\alpha Z)^3/Z^3$. The total values are compared with the ones from Refs.~\cite{volotka:2008:062507,shabaev:1998:149}.}
\label{table:final_results}
\centering
\tabcolsep5pt
\begin{tabular}{cllllll}
\hline\hline
Ion	& $X_{a}^{(0)}$ & $X_{a}^{(1)}$ & $X_{a}^{(2)}$ & $X_{a}^{(3+)}$
      & Total            & $\epsilon_{\rm sph}$ \\
\hline
$^{15}$N$^{4+}$	   & 0.618\,794\,6 & 0.019\,646\,3(2) & 0.003\,199\,9(3) &-0.000\,144(5)      
      & 0.641\,497(5)    & 0.000\,268\,0 \\
&&&&  & 0.641\,45$^{a}$& \\
$^{23}$Na$^{8+}$   & 0.758\,442\,6 & 0.015\,527\,7(1) & 0.001\,371\,5(2) &-0.000\,037\,0(13)
      & 0.775\,304\,9(14) & 0.000\,496\,5 \\
&&&&  & 0.775\,28$^{a}$& \\ 
$^{31}$P$^{12+}$   & 0.833\,026\,7 & 0.012\,682\,5    & 0.000\,754\,8    &-0.000\,014\,4(5) 
      & 0.846\,449\,6(6) & 0.000\,733\,5 \\
&&&&  & 0.846\,43$^{a}$& \\ 
$^{39}$K$^{16+}$   & 0.884\,885\,5 & 0.010\,832\,6    & 0.000\,476\,9    &-0.000\,007\,2(3) 
      & 0.896\,187\,8(5) & 0.001\,034\,0 \\
&&&&  & 0.896\,17$^{a}$& \\ 
$^{51}$V$^{20+}$   & 0.927\,883\,1 & 0.009\,600\,0    & 0.000\,329\,6    &-0.000\,003\,9(2)
      & 0.937\,808\,8(4) & 0.001\,355\,1 \\
&&&&  & 0.937\,81$^{a}$& \\ 
$^{55}$Mn$^{22+}$   & 0.948\,019\,4 & 0.009\,141\,2    & 0.000\,280\,9    &-0.000\,002\,9(1)
      &0.957\,438\,5(4)  & 0.001\,554\,7  \\
&&&&  & 0.957\,43$^{a}$& \\ 
$^{57}$Fe$^{23+}$   &  0.957\,971\,8 &  0.008\,942\,9    & 0.000\,260\,6  &-0.000\,002\,5(1)
      &0.967\,173\,1(4)  &0.001\,655\,7  \\
&&&&  & 0.967\,16$^{a}$& \\ 
$^{59}$Co$^{24+}$  & 0.967\,921\,1 & 0.008\,762\,6    & 0.000\,242\,5    &-0.000\,002\,3(1)
      & 0.976\,923\,9(4) & 0.001\,753\,4 \\
$^{61}$Ni$^{25+}$   &  0.977\,891\,1 & 0.008\,599\,4    & 0.000\,226\,4  &-0.000\,002\,0(1) 
      &0.986\,714\,5(4)  & 0.001\,859\,6 \\
&&&&  & 0.986\,71$^{a}$& \\ 
$^{69}$Ga$^{28+}$  & 1.008\,153\,0 & 0.008\,195\,4    & 0.000\,187\,1    &-0.000\,001\,4(1)
      & 1.016\,534\,1(4) & 0.002\,230\,8 \\
$^{79}$Br$^{32+}$  & 1.050\,690\,7 & 0.007\,824\,2    & 0.000\,149\,7    &-0.000\,001\,0
      & 1.058\,663\,6(4) & 0.002\,776\,1 \\
$^{89}$Y$^{36+}$   & 1.097\,198\,8 & 0.007\,605\,1    & 0.000\,124\,2    &-0.000\,000\,6
      & 1.104\,927\,5(4) & 0.003\,352\,3 \\
\hline\hline
\end{tabular}
\\[2mm]
\footnotesize{$^{ a}$ Ref.~\cite{volotka:2008:062507}, recalculated to the nuclear models and nuclear parameters employed in this paper.}\\
\footnotesize{$^{ b}$ Ref.~\cite{shabaev:1998:149}, recalculated to the nuclear models and nuclear parameters employed in this paper.}\\
\end{table}
\begin{table}
TABLE~\ref{table:final_results}. (\textit{Continued.})
\centering
\tabcolsep5pt
\begin{tabular}{cllllll}
\hline\hline
Ion	& $X_{a}^{(0)}$ & $X_{a}^{(1)}$ & $X_{a}^{(2)}$ & $X_{a}^{(3+)}$
      & Total            & $\epsilon_{\rm sph}$ \\
\hline
$^{98}$Tc$^{40+}$  & 1.148\,599\,7 & 0.007\,506\,6    & 0.000\,106\,1    &-0.000\,000\,5
      & 1.156\,212\,0(4) & 0.004\,094\,4 \\
$^{109}$Ag$^{44+}$ & 1.206\,400\,7 & 0.007\,514\,4    & 0.000\,093\,3    &-0.000\,000\,3
      & 1.214\,008\,1(4) & 0.004\,961\,0 \\
$^{121}$Sb$^{48+}$ & 1.272\,127\,8 & 0.007\,619\,3    & 0.000\,084\,3    &-0.000\,000\,3
      & 1.279\,831\,1(4) & 0.005\,937\,6 \\
&&&&  & 1.279\,5$^{b}$  & \\ 
$^{133}$Cs$^{52+}$ & 1.347\,184\,0 & 0.007\,816\,7    & 0.000\,078\,2    &-0.000\,000\,2
      & 1.355\,078\,7(4) & 0.007\,086\,5 \\
$^{141}$Pr$^{56+}$ & 1.433\,721\,5 & 0.008\,109\,6    & 0.000\,074\,7    &-0.000\,000\,2
      & 1.441\,905\,6(4) & 0.008\,410\,4 \\
&&&&  & 1.441\,6$^{b}$  & \\ 
$^{151}$Eu$^{60+}$ & 1.533\,117\,2 & 0.008\,497\,9    & 0.000\,073\,3    &-0.000\,000\,2
      & 1.541\,688\,2(4) & 0.010\,032\,4 \\
&&&&  & 1.541\,4$^{b}$  & \\ 
$^{165}$Ho$^{64+}$ & 1.648\,631\,7 & 0.008\,994\,7    & 0.000\,074\,0    &-0.000\,000\,2
      & 1.657\,700\,2(4) & 0.011\,958\,0 \\
&&&&  & 1.657\,4$^{b}$  & \\ 
$^{175}$Lu$^{68+}$ & 1.783\,175\,6 & 0.009\,611\,4    & 0.000\,076\,5    &-0.000\,000\,2
      & 1.792\,863\,3(4) & 0.014\,224\,2 \\
&&&&  & 1.792\,6$^{b}$  & \\ 
$^{185}$Re$^{72+}$ & 1.945\,250\,8 & 0.010\,390\,4    & 0.000\,081\,6    &-0.000\,000\,2
      & 1.955\,722\,6(4) & 0.016\,444\,1 \\
&&&&  & 1.955\,4$^{b}$  & \\ 
$^{197}$Au$^{76+}$ & 2.134\,320\,5 & 0.011\,326\,4    & 0.000\,089\,0    &-0.000\,000\,3
      & 2.145\,735\,7(4) & 0.019\,382\,4 \\
$^{209}$Pb$^{79+}$ & 2.300\,174\,6 & 0.012\,164\,4    & 0.000\,096\,6    &-0.000\,000\,4	
      &	2.312\,435\,2(4)	             & 0.021\,811\,3 \\
&&&&  & 2.311\,8$^{b}$  & \\ 
%$^{209}$Bi$^{80+}$ & 2.360\,391\,4 & 0.012\,471\,3    & 0.000\,099\,5(1)    &-0.000\,000\,4
%      &	2.372\,961\,8(4) & 0.022\,694\,9 \\
%&&&&  & 2.372\,3$^{b}$  & \\
%&&&&  & 2.372\,959(9)$^{c}$  & \\
\hline\hline
\end{tabular}
\\[2mm]
\footnotesize{$^{ b}$ Ref.~\cite{shabaev:1998:149}, recalculated to the nuclear models and nuclear parameters employed in this paper.}\\
\end{table}
%
% ============================= CONCLUSION ==============================
%
\section{CONCLUSION}
To summarize, we evaluate the interelectronic-interaction contribution to the ground-state hyperfine splitting in Li-like ions for the wide range of the nuclear charge numbers. The contributions due to the one- and two-photon-exchange corrections are treated within the rigorous QED approach in the framework of the extended Furry picture. The higher-order interelectronic-interaction terms were taken into account by means of the recursive perturbation theory. As a result, we substantially improve the accuracy of the interelectronic-interaction corrections to the ground-state hyperfine splitting in Li-like ions in the range $Z = 7 - 82$.
These calculations represent also an important prerequisite for the evaluation of the specific difference between H- and Li-like ions which can serve for high-precision tests of the bound-state QED in strong nuclear field.
 In order to push forward the test of QED with hfs we plan to evaluate the screened QED corrections to the hfs in Li-like ions for a wide range of the nuclear charge $Z$. These results, combined with the present rigorous calculations of the interelectronic-interaction correction, would allow us to construct the specific differences, where the QED effects can be tested by comparison with experiment.

%
% ======================== ACKNOWLEDGEMENTS ============================
%%=======================================================================
\section{Acknowledgments}
This work was supported by DFG (VO1707/1-3) and RFBR (Grant No. 19-02-00974). D.A.G. acknowledges the support by the Foundation for the Advancement of Theoretical Physics and Mathematics ``BASIS''.
\section{APPENDIX A: Two-photon exchange counterterm contribution}
The formal expressions for the contribution of the counterterm diagrams depicted in Fig.~\ref{fig-countertems} are given by
\be
X^{(2)}_{a} = X^{(2)\rm -ct-1}_{a} + X^{(2)\rm -ct-2}_{a}
\ee
where
\be
\label{Atype}
X^{(2)\rm -ct-1}_{a} &=& 2\,G_a \sum_b \Bigg[                   
   \la \xi_a \eta_b \vert I(0) \vert ab \ra - \la \eta_b \xi_a \vert I(\Delta) \vert ab \ra
 + \la \eta_a \xi_b \vert I(0) \vert ab \ra - \la \xi_b \eta_a \vert I(\Delta) \vert ab \ra
   \nonumber\\
&+&\la \xi_a b \vert I(0) \vert \eta_a b\ra - \la b \xi_a \vert I(\Delta) \vert \eta_a b\ra
 + \la a \xi_b \vert I(0) \vert a \eta_b\ra - \la \xi_b a \vert I(\Delta) \vert a \eta_b\ra 
   \nonumber\\
&+&\la \xi_a b \vert I(0) \vert a \eta_b\ra - \la \xi_b a \vert I(\Delta) \vert \eta_a b\ra
 + \la \eta_a b \vert I(0) \vert a \xi_b\ra - \la \eta_b a \vert I(\Delta) \vert \xi_a b\ra 
   \nonumber\\
&+&\Bigl( \la ab \vert I(0) \vert ab \ra - \la ba \vert I(\Delta) \vert ab \ra \Bigr)      
   \Bigl( \la\xi'_a \vert V_{\rm scr}\vert a\ra + \la\xi'_b \vert V_{\rm scr}\vert b\ra \Bigr)
 - \la ba \vert I^\prime(\Delta) \vert ab \ra
   \nonumber\\
&\times&
   \Bigl(\la\xi_a \vert V_{\rm scr} \vert a \ra - \la\xi_b \vert V_{\rm scr} \vert b \ra\Bigr)
 - \la a \vert T_0 \vert a \ra \la \eta_b a \vert I^\prime(\Delta) \vert ab \ra
 + \la b \vert T_0 \vert b \ra \la b \eta_a \vert I^\prime(\Delta) \vert ab \ra
   \nonumber\\
&-&\la a \vert V_{\rm scr} \vert a \ra \la \xi_b a \vert I^\prime(\Delta) \vert ab \ra
 + \la b \vert V_{\rm scr} \vert b \ra \la b \xi_a \vert I^\prime(\Delta) \vert ab \ra
 - \frac12 \la ba \vert I^{\prime\prime}(\Delta) \vert ab \ra
   \nonumber\\
&\times&  
   \Bigl( \la a \vert T_0 \vert a \ra - \la b \vert T_0 \vert b \ra \Bigr)  
   \Bigl( \la a \vert V_{\rm scr} \vert a \ra - \la b \vert V_{\rm scr} \vert b \ra \Bigr)  
   \nonumber\\          
&+&(-1)^P \sum_P \Bigl(
   \la \xi_a \vert V_{\rm scr} \vert \zeta_{b\vert PaPb} \ra
 + \la \xi_b \vert V_{\rm scr} \vert \zeta_{a\vert PbPa} \ra
 + \la \eta_a \vert T_0 \vert \zeta_{b\vert PaPb} \ra
   \nonumber\\       
&+&\la \eta_b \vert T_0 \vert \zeta_{a\vert PbPa} \ra
 + \la a \vert T_0 \vert a \ra \la a \vert V_{\rm scr} \vert \zeta'_{b\vert PaPb} \ra
 + \la b \vert T_0 \vert b \ra \la b \vert V_{\rm scr} \vert \zeta'_{a\vert PbPa} \ra
   \nonumber\\ 
&+&\la a \vert V_{\rm scr} \vert a \ra \la a \vert T_0 \vert \zeta'_{b\vert PaPb} \ra
 + \la b \vert V_{\rm scr} \vert b \ra \la b \vert T_0 \vert \zeta'_{a\vert PbPa} \ra
                 \Bigr)               \Bigg]
\ee
corresponds to the five diagrams from the upper part of Fig.~\ref{fig-countertems}, and
\be
X^{(2)\rm -ct-2}_{a} = G_a \Bigl( 2\la \xi_a \vert V_{\rm scr} \vert \eta_a \ra
  + 2 \la \xi'_a \vert V_{\rm scr} \vert a \ra \la a \vert V_{\rm scr} \vert a \ra
  + \la \eta_a \vert T_0 \vert \eta_a \ra
  + \la \eta'_a\vert V_{\rm scr} \vert a \ra \la a \vert T_0\vert a \ra \Bigr)
\ee
stays for the two diagrams from the lower part of Fig.~\ref{fig-countertems}. The employed functions are defined by Eqs.~\eqref{eq:eta} and \eqref{eq:zeta} together with
\be
| \xi_a \ra = {\sum_n}' \frac{| n \ra \la n | T_0 | a \ra}{\veps_a-\veps_n}\,,\;\;\;
| \xi^\pr_a \ra = \frac{\partial}{\partial\veps_a} | \xi_a \ra
\ee
and
\be
| \eta^\pr_a \ra = \frac{\partial}{\partial\veps_a} | \eta_a \ra\,,\;\;\;
| \zeta^\pr_{b|PaPb} \ra = \frac{\partial}{\partial\veps_a} | \zeta_{b|PaPb} \ra\,,\;\;\;
| \zeta^\pr_{a|PbPa} \ra = \frac{\partial}{\partial\veps_b} | \zeta_{a|PbPa} \ra\,.
\ee
The sum over $b$ runs over $1s$ states with $m_b = \pm 1/2$ , $\Delta = \veps_a - \veps_b$, and $I^{\prime\prime}(\omega) = d^2 I(\omega) / d \omega^2$.
%
% ======================================================================
%
%\bibliographystyle{apsrev4-1}
%\bibliography{liter}
\end{document}